# How Proposal Novelty, Topical Diversity, and Theory-Practice Balance Shape Scholarly Outcomes in Funded Education Research

Yunfeng Gao [1,†], Yuxuan Xiao [2,†], Jiaming Zhang [3], Yang Ding [4,*]

[1] School of Economics and Management, East China Normal University, Shanghai 200062, China

[2] Department of Education, Amity Global Institute, Singapore 238853, Singapore

[3] School of Mathematics and Statistics, The University of Glasgow, Glasgow G12 8TA, United Kingdom

[4] Business School, The University of Edinburgh, Edinburgh EH8 9JS, United Kingdom

[†] Yunfeng Gao and Yuxuan Xiao contributed equally to this manuscript.

* Corresponding Author: Yang Ding (yang.ding@ed.ac.uk)

**Abstract**

Education research occupies a distinctive position in public science because it is expected to advance scholarly knowledge while also informing learning, teaching, participation, and workforce development. This study examines how the intellectual characteristics of NSF-funded education proposals are associated with the subsequent academic performance of funded scholars. Linking 8,715 NSF education awards from 1990 to 2020 with 84,519 publications by principal investigators, the analysis focuses on four major NSF education divisions that collectively span undergraduate and graduate levels, formal and informal learning environments, and inclusive educational initiatives. Proposal novelty is measured as semantic distance from prior funded projects within the same division, topical diversity as breadth across latent research themes, and intellectual orientation as theoretical, practical, or balanced. The results show that NSF education funding is consistently associated with higher publication output across divisions. However, this increase is not accompanied by stronger citation performance or higher journal-level visibility; citation and CiteScore estimates are often negative, particularly in later decades. Proposal novelty shows limited and uneven associations with post-award outcomes, whereas topical diversity is more clearly related to publication growth in some divisions but weaker citation-based performance in others. Balanced proposals that integrate theoretical and practical aims display the most favourable overall profile, combining positive publication associations with fewer negative citation-based patterns. These findings highlight the importance of evaluating education research funding through multiple academic outcomes and division-specific research contexts.

# 1. Introduction

Public research funding is commonly assessed through the scholarly activity it enables, the knowledge trajectories it supports, and the visibility of the outputs produced by funded investigators. In education research, this assessment problem is particularly important because federal investments are expected to strengthen the knowledge base for learning, teaching, participation, and workforce development while also sustaining the academic communities that produce such knowledge. The National Research Council's account of scientific research in education emphasises systematic inquiry, theoretical grounding, empirical evidence, and relevance to educational problems (National Research Council, 2002). Current STEM education programmes supported by the United States National Science Foundation (NSF) adopt a related position. The EDU Core Research programme supports fundamental research on STEM learning and learning environments, broadening participation, and STEM workforce development (National Science Foundation, 2026a). NSF's FY 2026-2030 Strategic Plan also places STEM education and workforce development within the agency's wider mission of research, innovation, infrastructure, and national scientific capacity (National Science Foundation, 2026b). These priorities make NSF-funded education research a useful setting for examining how public funding is associated with the academic performance of funded scholars.

This study focuses on three observable dimensions of academic performance: publication output, citation performance, and journal-level visibility. These indicators do not exhaust the value of education research, but they remain central to the organisation and evaluation of academic science. Publications are the principal channel through which research claims enter cumulative scholarly debate. Citations provide one measure of subsequent uptake within academic literatures. Journal-level indicators, such as CiteScore, approximate the visibility of publication venues within indexed scholarly communication. The responsible-metrics literature cautions that such indicators should be interpreted in relation to field, purpose, and evaluation context rather than treated as direct measures of intrinsic research quality (Hicks et al., 2015). This caution is especially relevant for education research, where scholarly outputs may differ across programme missions, audiences, and time horizons. It does not make bibliometric outcomes irrelevant; rather, it requires analysing them carefully and separately.

A substantial body of research examines whether grant funding is associated with subsequent scientific productivity. Jacob and Lefgren (2011) study the impact of research grant funding on scientific output, while Ding and Moreira (2025) use NSF social science and economics data to show that estimates of funding effects are sensitive to selection, timing, and model specification. This literature suggests that the relationship between funding and academic performance should not be assumed to be automatic or homogeneous. Research grants may provide resources, time, personnel, data access, and institutional support, but publication and citation outcomes also depend on prior researcher trajectories, field structure, topic maturity, collaboration patterns, and the audiences reached by funded work. For NSF education research, these issues are intensified by the heterogeneity of programme divisions and by the diversity of proposal aims.

The present study therefore shifts attention from whether funding is associated with academic performance in general to how the textual characteristics of funded proposals are related to post-award academic trajectories. Grant proposals are not merely administrative documents. They encode research problems, conceptual orientations, methodological commitments, intended audiences, and expected contributions. In the NSF education portfolio, proposals may emphasise explanatory theory, instructional intervention, assessment design, institutional change, broadening participation, or combinations of these purposes. Treating funded projects as a homogeneous category may therefore obscure meaningful variation in the kinds of academic outputs that follow different forms of funded work.

Three proposal-level characteristics are central to the analysis: novelty, topical diversity, and theoretical-practical orientation. First, building on literature conceptualising new knowledge as atypical recombination (Uzzi et al., 2013) that may face delayed recognition (Wang et al., 2017), we operationalise proposal novelty as the semantic distance from previously funded projects within the same education division. Second, because education research integrates multiple disciplines, we measure topical diversity, a dimension conceptually distinct from novelty, as the extent to which a proposal spans latent themes in the funding corpus, capturing its multidimensional intellectual breadth (Stirling, 2007). Finally, we assess the proposal's theoretical-practical orientation. Recognising that productive education research often crosses boundaries between academic inquiry and applied practice (Penuel et al., 2015), we classify projects as theoretical, practical, or balanced based on whether they emphasise conceptual development, educational implementation, or the explicit integration of both.

Empirically, the study links 8,715 NSF-funded education projects from 1990 to 2020 with 84,519 scholarly publications produced by principal investigators. The analysis focuses on four major NSF education divisions: Graduate Education (DGE), Research on Learning (DRL), Undergraduate Education (DUE), and Human Resource Development (HRD). These divisions are analysed separately because they differ in programme missions, funded populations, and expected forms of scholarly output. Proposal novelty is measured through semantic distance from earlier funded proposals within the same division. Topical diversity is measured using topic-distribution entropy. Theoretical, practical, and balanced orientations are identified through a structured text-based classification procedure.

The empirical strategy uses a staggered funding-timing design. Rather than comparing funded scholars with permanently unfunded scholars, the analysis compares scholars after their first observed NSF education award with not-yet-funded scholars who later receive NSF education awards. This design places the comparison within the population of eventual NSF education awardees and is better aligned with recent research on grant timing and academic productivity. The outcomes are measured annually at the scholar-year level and include publication counts, average citations per publication, and average journal CiteScore.

The study contributes to research on science policy and education research funding in three ways. First, it provides large-scale evidence on the academic trajectories of NSF-funded education researchers, a population that has received less systematic attention than researchers in biomedical, natural science, or general social science funding studies. Second, it moves beyond award receipt alone by analysing how proposal novelty, topical diversity, and theoretical-practical orientation are associated with subsequent academic performance. Third, it treats publication output, citation performance, and journal-level visibility as distinct outcomes, allowing the analysis to examine whether the same funded projects follow similar or different trajectories across these dimensions of academic performance.

# 2. Theoretical Background

## 2.1 Use-Inspired Education Research and the Evaluation of Public Funding

This study is situated within the tradition of use-inspired research. Stokes's account of Pasteur's Quadrant challenged the simple distinction between basic and applied science by arguing that some research is simultaneously motivated by the pursuit of fundamental understanding and by consideration of practical use (Stokes, 1997). This framework is particularly relevant for education research. Education projects often seek to contribute to generalisable knowledge about learning, teaching, assessment, equity, or institutional change, while also addressing practical problems faced by educators, students, institutions, and policy systems.

The use-inspired framework provides a more suitable theoretical basis for this study than a narrow productivity model. Public research funding is often evaluated through publications, citations, and journal indicators, yet these indicators capture only part of the value of mission-oriented research. In education, scholarly outputs may be accompanied by curriculum materials, instructional designs, assessment tools, professional development models, institutional partnerships, or policy-relevant evidence. The National Research Council's account of scientific research in education emphasises systematic inquiry, theory, evidence, and relevance to educational problems, rather than reducing research quality to publication volume alone (National Research Council, 2002). Similarly, work on research use in education shows that evidence may be used instrumentally, conceptually, politically, or through longer processes of sense-making and organisational learning (Farley-Ripple et al., 2020; Tseng, 2012)

This theoretical position is important for interpreting the four NSF education divisions examined in this study. DGE, DRL, DUE, and HRD do not represent identical funding populations. DGE is closely connected to graduate education and researcher development. DRL is more directly associated with learning research, assessment, and educational innovation. DUE is centred on undergraduate STEM education, curriculum improvement, and instructional reform. HRD is more closely associated with broadening participation, equity, and institutional capacity. These divisions therefore occupy different positions within the broader use-inspired research space. Their scholarly outputs may differ not only in quantity but also in intended audience, dissemination channel, and relationship to practice.

Research-practice partnership scholarship further clarifies why education research should not be evaluated only through academic visibility. Penuel et al. (2015) describe partnership work as joint activity across the boundaries of research and practice rather than a simple transfer of findings from researchers to practitioners. Coburn et al. (2015) show that such partnerships are often organised around persistent problems of practice, shared authority, and long-term collaboration. These features are relevant to NSF education programmes, especially where projects involve interventions, institutional improvement, or participation-oriented goals. They suggest that education research funding may be associated with academic outputs, but those outputs are only one expression of a broader knowledge-production process.

The present study therefore treats publications, citations, and CiteScore as indicators of academic performance, not as exhaustive measures of research value. This distinction is central. A positive association with publication output may indicate increased scholarly activity, whereas weaker associations with citations or journal-level indicators may reflect differences in audience, field structure, applied orientation, or time needed for diffusion. The theoretical expectation is not that NSF education funding should uniformly increase every bibliometric indicator. Rather, the use-inspired framework motivates a more differentiated analysis of how academic performance varies by division, proposal novelty, topical diversity, and theoretical-practical orientation.

## 2.2 Novelty, Topical Diversity, and the Uneven Rewards of Intellectual Breadth

Scientific novelty and diversity are central concepts in the science of science literature, but they are usually measured through specific empirical designs rather than treated as abstract virtues. A common starting point is the view that new knowledge is produced by recombining existing ideas, methods, objects, or fields in ways that depart from established patterns. Uzzi et al. (2013) operationalise novelty through atypical combinations of journals in reference lists, showing that influential papers often combine conventional intellectual foundations with unusual pairings. Foster et al. (2015) use networks of biomedical concepts to distinguish strategies that deepen established knowledge from those that introduce new entities or relationships. Wang et al. (2017) measure novelty as first-time combinations of referenced journals and show that highly novel papers may generate large long-run gains while facing delayed recognition and greater citation uncertainty.

These studies are important because they make novelty observable, but they also reveal that novelty measures are design-dependent. Citation-based approaches generally infer novelty from the knowledge base that a published article cites. They therefore capture realised scholarly products after authors have selected references, journals have imposed genre conventions, and audiences have begun to evaluate the work. In grant-proposal settings, however, the empirical object is different. A proposal describes an intended research agenda before the project has produced articles, reference lists, or citation traces. Measuring proposal novelty therefore requires comparing the semantic content of proposed work with the prior corpus of funded proposals, rather than relying on the backward citations of later publications.

Beyond novelty, researchers have also investigated the roles of knowledge diversity, interdisciplinarity, and intellectual breadth. Stirling (2007) provides the widely used distinction among variety, balance, and disparity: diversity increases when work draws on more categories, distributes attention more evenly across them, and spans more distant knowledge domains. Porter and Ràfols (2009) and Rafols and Meyer (2009) translate this framework into science-mapping and bibliometric indicators, often using disciplinary categories, citation distributions, and Rao-Stirling diversity. Yegros-Yegros et al. (2015) further show that these dimensions can have different and sometimes nonlinear relationships with citation impact: variety may be beneficial, whereas excessive balance or disparity can make work harder for audiences to place.

The funding and evaluation literature also cautions that intellectual breadth is not always rewarded evenly. Bromham et al. (2016) find that interdisciplinary proposals have lower funding success in Australian Research Council data, and Boudreau et al. (2016) show experimentally that novelty and intellectual distance can shape proposal evaluation. Fontana et al. (2020) further caution that indicators of novelty and interdisciplinarity may overlap when both are derived from unusual combinations of cited knowledge domains. Taken together, this literature suggests that novelty and diversity are analytically related but should not be collapsed into a single measure.

This distinction is especially important for NSF-funded education research. Education proposals often combine learning science, psychology, sociology, computer science, assessment, policy, institutional research, and equity-oriented scholarship. They may also address multiple audiences, including researchers, teachers, institutions, programme designers, and policy actors. As a result, intellectual breadth may be substantively valuable even when it does not translate into immediate citation advantage. A topically broad education project may generate curriculum tools, implementation knowledge, partnership infrastructure, or equity-relevant evidence whose academic visibility differs from that of a narrower disciplinary article.

NSF education proposals also differ from the publication corpora on which many novelty and interdisciplinarity indicators were developed. First, proposals are prospective documents: they state aims, contexts, methods, and expected contributions before publication outcomes occur. Second, the relevant comparison set is institutional and programmatic. DGE, DRL, DUE, and HRD have different histories, missions, and intellectual communities, so novelty should be judged relative to the prior trajectory of each division rather than against a single pooled NSF education corpus. Third, proposal abstracts usually do not contain stable reference lists or field classifications, making citation-category measures less suitable than text-based semantic and topic-distribution measures.

We measure proposal novelty as semantic distance from earlier funded proposals within the same NSF education division, capturing how far a project departs from the division-specific textual history of funded work. We measure topical diversity as the entropy of a proposal's topic distribution, capturing how broadly its abstract spans latent themes in the NSF education corpus. The two measures are therefore designed to reflect the proposal-stage and division-specific nature of the data while preserving the central insight of the broader literature: novelty concerns departure from prior knowledge patterns, whereas topical diversity concerns the breadth of knowledge domains combined within a project.

### 2.3 Theoretical, Practical, and Balanced Orientations in Education Research

The use-inspired framework does not treat theory and practice as opposing categories. Instead, it allows research to vary along two dimensions: the pursuit of fundamental understanding and the consideration of practical use (Stokes, 1997). This is well suited to education research, where projects may develop conceptual models of learning, design instructional interventions, evaluate programmes, or integrate theory with implementation.

The education research literature has long recognised this dual responsibility. Scientific research in education is expected to contribute to reliable knowledge while engaging with educational problems that matter for learners, educators, and institutions (National Research Council, 2002). Research-use scholarship further shows that the movement from research to practice is rarely linear. Evidence must be interpreted, translated, adapted, and negotiated within specific organisational contexts (Farrell et al., 2019). These studies suggest that education research with practical relevance may not always maximise conventional citation indicators, particularly when its primary audiences include educators, programme leaders, or policy actors.

At the same time, purely practical orientation should not be assumed to be academically weaker, nor should theoretical orientation be assumed to be academically stronger. Practical projects may produce multiple outputs because interventions, evaluations, and implementation studies generate publishable evidence across settings. Theoretical projects may have slower diffusion but contribute to conceptual development. Balanced projects may be positioned to connect scholarly and applied audiences, but their performance remains an empirical question rather than a theoretical certainty. The study therefore classifies proposals as theoretical, practical, or balanced to examine whether these orientations are associated with different academic trajectories.

This classification also aligns with the institutional heterogeneity of DGE, DRL, DUE, and HRD. DRL may contain more projects that connect learning theory, assessment, and educational design. DUE may include a larger share of curriculum and instructional improvement work. HRD may include projects whose practical relevance is tied to participation, equity, and institutional capacity. DGE may include projects that link graduate training with professional development and research capacity. A theory-practice orientation variable allows the analysis to examine whether these differences are reflected in scholarly output patterns without imposing a single hierarchy of value across divisions.

# 3. Data

## 3.1 NSF Education Research Funding

The award dataset contains 8,715 NSF education projects funded between 1990 and 2020. The sample is restricted to awards administered through DGE, DRL, DUE, and HRD. Divisions with very small numbers of awards, unstable administrative boundaries, or substantial reorganisation over the observation period are excluded to improve comparability across years.

The award-level records include the NSF award number, project title, abstract, start year, division, funding amount, award mechanism, and principal investigator information. The project title and abstract are used to construct proposal-level measures of novelty, topical diversity, and theoretical-practical orientation. The award start year is used to define the timing of each scholar's first observed NSF education award within the relevant division.

The sample is limited to scholars affiliated with U.S. universities and comparable higher education institutions. Awards to industry organisations, K-12 schools, and non-academic organisations are excluded because their publication incentives and output profiles are unlikely to be directly comparable with those of university-based researchers. The analysis focuses on Standard Grants and Continuing Grants, while excluding fellowships, contracts, and other award types whose purposes and expected outputs differ from research project grants. To prevent potential spillover effects from multiple NSF awards on academic performance, our study excluded scholars who received two or more NSF grants.

## 3.2 Publication and Bibliometric Data

Publication data are constructed by linking NSF award records to bibliometric records from Web of Science and Microsoft Academic Graph. Web of Science is used as the initial bridge because its funding acknowledgement fields contain NSF award numbers for a subset of publications. These award-number links provide a conservative way to identify publications explicitly connected to NSF-funded projects.

The matched Web of Science records are then linked to Microsoft Academic Graph primarily through DOI identifiers. MAG is used to recover broader publication portfolios for the principal investigators associated with NSF education awards. This procedure provides a longitudinal record of scholarly outputs before and after the scholar's first observed NSF education award.

The final funded-scholar publication file contains 84,519 publications authored by NSF-funded principal investigators. These records include publication titles, publication years, document types, citation counts, research field tags, author identifiers, and institutional information where available. Detailed procedures regarding the integration of funding records with bibliometric databases and the subsequent sample retention criteria are provided in **Supplementary Note 1**.

The main outcomes are measured annually at the scholar-year level. Publication output is defined as the number of publications associated with a scholar in a given year. Citation performance is measured as the average citation count of that scholar's publications in a given year. Journal-level visibility is measured using average journal CiteScore, matched to the journals in which the scholar published. CiteScore is treated as a journal-level indicator of citation visibility rather than as a direct measure of article quality.

### 3.3 Analytical Panel and Not-Yet-Funded Comparison Group

The empirical analysis is based on a longitudinal scholar-year panel. Importantly, the comparison group is not composed of scholars who never received NSF education funding. Instead, all scholars in the analytical sample eventually received NSF education funding. The comparison is based on variation in the timing of first award receipt.

For each scholar, the first observed NSF education award within DGE, DRL, DUE, or HRD defines the first award year. Scholars who have already received their first award contribute post-award observations. Scholars who will receive an award later, but have not yet received one in a given year, serve as not-yet-funded comparison scholars during their pre-award period. Thus, the control group is composed of future NSF awardees who are observed before their own first award.

This design follows the logic of staggered funding-timing studies, including research on grant receipt and productivity in the social sciences and economics. It is preferable to comparing funded scholars with never-funded scholars because eventual awardees are more likely to belong to comparable research communities and to share similar grant-seeking orientations. The design therefore compares earlier-funded scholars with later-funded scholars rather than funded scholars with researchers permanently outside the NSF education funding system.

The comparison group changes over time. For example, a scholar first funded in 2010 may serve as a not-yet-funded comparison scholar for a scholar first funded in 2005 during the years before 2010. After receiving the first award, that scholar enters the post-award group and no longer contributes untreated comparison observations. The resulting design uses the staggered timing of first award receipt to compare post-award academic trajectories with the contemporaneous trajectories of scholars who had not yet received their first award.

# 4. Methods

## 4.1 Measuring Proposal Novelty

Proposal novelty is measured as the semantic distance between a focal proposal and prior proposals funded within the same NSF education division. The title and abstract of each proposal are concatenated and converted into sentence-level embeddings using the SciBERT transformer model. This model represents each proposal as a dense semantic vector and allows proposals to be compared by contextual meaning rather than by surface-level word overlap.

For each proposal $i$ funded in year $t$ within division $d$, the reference set consists of proposals funded in the same division before year $t$. Novelty is calculated as one minus the average cosine similarity between the focal proposal and all prior proposals in that divisional history:

$$Novelty_{idt} = 1 - \frac{1}{N_{d,t-1}} \sum_{j \in d, \tau_j < t} cosine(e_i, e_j) \quad (1)$$

where $e_i$ is the embedding vector for proposal $i$, $j$ is the embedding vector for prior proposal $j$, and $N_{d,t-1}$ is the number of prior proposals in division $d$ before year $t$. Higher values indicate greater semantic distance from earlier funded work in the same division.

4.2 Measuring Topical Diversity

Topical diversity is measured using SciBERT, a topic modelling framework that combines transformer-based document embeddings, dimensionality reduction, clustering, and class-based TF-IDF representations (Beltagy et al., 2019; McInnes et al., 2018; Saputra et al., 2020). SciBERT is applied to the NSF education proposal corpus to identify latent thematic areas across project titles and abstracts.

For each proposal, the model provides a distribution across identified topics. Topical diversity is calculated using Shannon entropy:

$$Diversity_i = -\sum_{k=1}^{K} p_{i,k} \ln(p_{i,k}) \quad (2)$$

where $p_{i,k}$ is the estimated probability that proposal $i$ is associated with topic $k$, and $K$ is the number of topics identified in the proposal corpus. Higher entropy indicates that a proposal is distributed more evenly across multiple topics, whereas lower entropy indicates concentration in a narrower thematic area.

This measure captures thematic breadth rather than research quality. A highly diverse proposal is not assumed to be more rigorous, more innovative, or more valuable than a specialised proposal. It is simply more broadly distributed across the estimated topic structure of NSF education research. Topical diversity scores are standardised before regression analysis.

we compared the distributions of micro-level proposal diversity and macro-level topic coverage across different divisions (see **Supplementary Note 2**). The observed significant variations among divisions strongly indicate that treating all education-related funded projects as a homogeneous entity may obscure underlying structural characteristics. Consequently, it is necessary to adopt a more granular, division-level analytical framework in our subsequent empirical design to more accurately capture the relationship between proposal characteristics and academic performance within specific funding contexts.

### 4.3 Classifying Theoretical, Practical, and Balanced Orientations

Each proposal abstract is classified into one of three intellectual orientations: theoretical, practical, or balanced. The classification is based on the substantive emphasis of the proposal abstract rather than on keywords alone. A proposal is classified as theoretical when its primary emphasis is on conceptual development, explanatory models, methodological innovation, or generalisable understanding of learning, teaching, or educational systems. A proposal is classified as practical when its primary emphasis is on instructional design, intervention, implementation, programme development, evaluation, or direct improvement of educational practice. A proposal is classified as balanced when it explicitly combines theoretical development with practical application, for example by using theory to design an intervention or by using implementation evidence to refine conceptual understanding.

The classification is implemented using GPT-4-assisted semantic coding. The model is prompted with a structured coding rubric and asked to classify each proposal abstract according to the definitions above. Low-confidence classifications are excluded or reviewed before the final orientation variable is constructed. The resulting labels are used as proposal-level categorical indicators in empirical analysis. To systematically categorize the textual emphasis of each project, we developed an assisted classification procedure that is fully documented in **Supplementary Note 3**.

4.4 Staggered Funding-Timing Specification

The main empirical analysis uses a staggered-adoption difference-in-differences-style specification. The design exploits variation in the timing of first NSF education funding among scholars who all eventually receive an award. At any given point in time, scholars who have already received their first NSF education award are compared with scholars who will receive an award later but have not yet received one.

The baseline model is:

$$Y_{i,t} = \alpha_i + \lambda_t + \beta PostAward_{i,t} + \boldsymbol{\gamma X_{i,t'}} + \varepsilon_{i,t} \tag{3}$$

where $Y_{i,t}$ denotes the academic outcome of scholar $i$ in year $t$. The outcomes include annual publication count, average citations per publication, and average journal CiteScore. $PostAward_{i,t}$ equals one for years after scholar $i$'s first observed NSF education award and zero for years before that award. Scholar fixed effects, $\alpha_i$, account for time-invariant differences across researchers. Year fixed effects, $\lambda_t$, account for common changes in publication practices, database coverage, citation environments, and broader academic conditions. $\boldsymbol{X_{i,t}}$ includes observable covariates. The coefficient $\beta$ describes the average post-award difference in academic outcomes for scholars after receiving NSF education funding, relative to their own pre-award outcomes and to the contemporaneous outcomes of not-yet-funded scholars.

To isolate the effect of the focal explanatory variables on scholars' academic performance, we control for a set of individual-, research-, and institution-level factors that may also influence research productivity. At the individual level, we control for scholars' academic age, gender, prior research performance, and funding amount. Academic age is measured as the number of years between a scholar's first publication and the year in which the focal funding was awarded, capturing differences in career stage and accumulated research experience. Prior research performance is measured using scholars' publication output, citation counts, and CiteScores during the first three years of their academic careers. Specifically, the logarithms of publication and citation counts are employed to reduce skewness. In addition, a dummy variable indicating this initial career period is included to account for path dependence in research performance and potential reverse-causality concerns. To account for disciplinary differences in publication and citation practices, we further control for the overall publication and citation volumes within each scholar's research field, classified according to the MAG field taxonomy. These variables capture field-specific norms and opportunities for knowledge production and impact. Finally, institutional factors are controlled for by including the publication and citation outputs of scholars' affiliated universities, which reflect institutional research capacity, publication orientation, and academic influence.

### 4.5 Treatment Timing and Comparison Structure

Let $G_i$ denote the first NSF education award year for scholar $i$ within the relevant division. A scholar is observed in the pre-award period when $t < G_i$ and in the post-award period when $t \geq G_i$. Scholars with later first-award years serve as not-yet-funded comparison scholars for those funded earlier, provided they have not yet entered their own post-award period. This design has two advantages. First, it avoids relying on never-funded scholars, who may differ substantially from NSF awardees in unobserved grant-seeking behaviour, institutional resources, research networks, or career orientation. Second, it aligns the comparison more closely with the NSF funding process by using scholars who eventually entered the same funding system. However, the design remains observational. Differences between earlier-funded and later-funded scholars may still reflect unobserved differences in academic maturity, prior momentum, or proposal readiness. The estimates should therefore be read as structured longitudinal associations.

To examine whether proposal characteristics are associated with different post-award academic trajectories, the baseline model is extended by interacting the post-award indicator with proposal novelty and topical diversity:

$$Y_{i,t} = \alpha_i + \lambda_t + \beta PostAward_{i,t} + \theta(PostAward_{i,t} \times Z_i) + \gamma \mathbf{X}_{i,t'} + \varepsilon_{i,t} \quad (4)$$

where $Z_i$ represents either standardised proposal novelty or standardised topical diversity. The coefficient $\theta$ indicates whether the estimated post-award association differs according to the textual characteristics of the funded proposal.

### 4.6 Robustness

To evaluate the empirical validity of our staggered estimation design, we conducted a series of diagnostic tests and sensitivity analyses. We first estimated a dynamic event study specification to systematically assess the parallel trends assumption underlying the main baseline models. This dynamic approach confirms that the academic trajectories of the scholars prior to receiving the funding are statistically indistinguishable from their unfunded counterparts, with the set of estimates available in **Supplementary Note 4**. Furthermore, we performed a placebo timing falsification test to examine whether the observed associations might be driven by unobserved temporal confounders. By artificially shifting the onset of the funding prior to the actual award, this test yields no statistically significant estimates, thereby reinforcing the stability of our core findings as thoroughly detailed in **Supplementary Note 5**.

## 5. Results

### 5.1 Descriptive Patterns in Proposal Novelty and Topical Diversity

Figure 1 describes the evolution of proposal novelty and topical diversity across the four NSF education divisions between 1990 and 2020. Figure 1(a) shows that mean proposal novelty does not increase monotonically over time. Instead, the most visible pattern is divisional heterogeneity. HRD starts from the highest level of novelty in the early 1990s, with values above 0.85, but then declines gradually and stabilises closer to 0.70 after the mid-2000s. DGE displays greater volatility: its novelty score is relatively high in the early 1990s, falls sharply around the mid-1990s, and remains mostly between 0.60 and 0.70 thereafter. By contrast, DRL and DUE follow more stable trajectories, with mean novelty concentrated around 0.65 to 0.70 for most of the period.

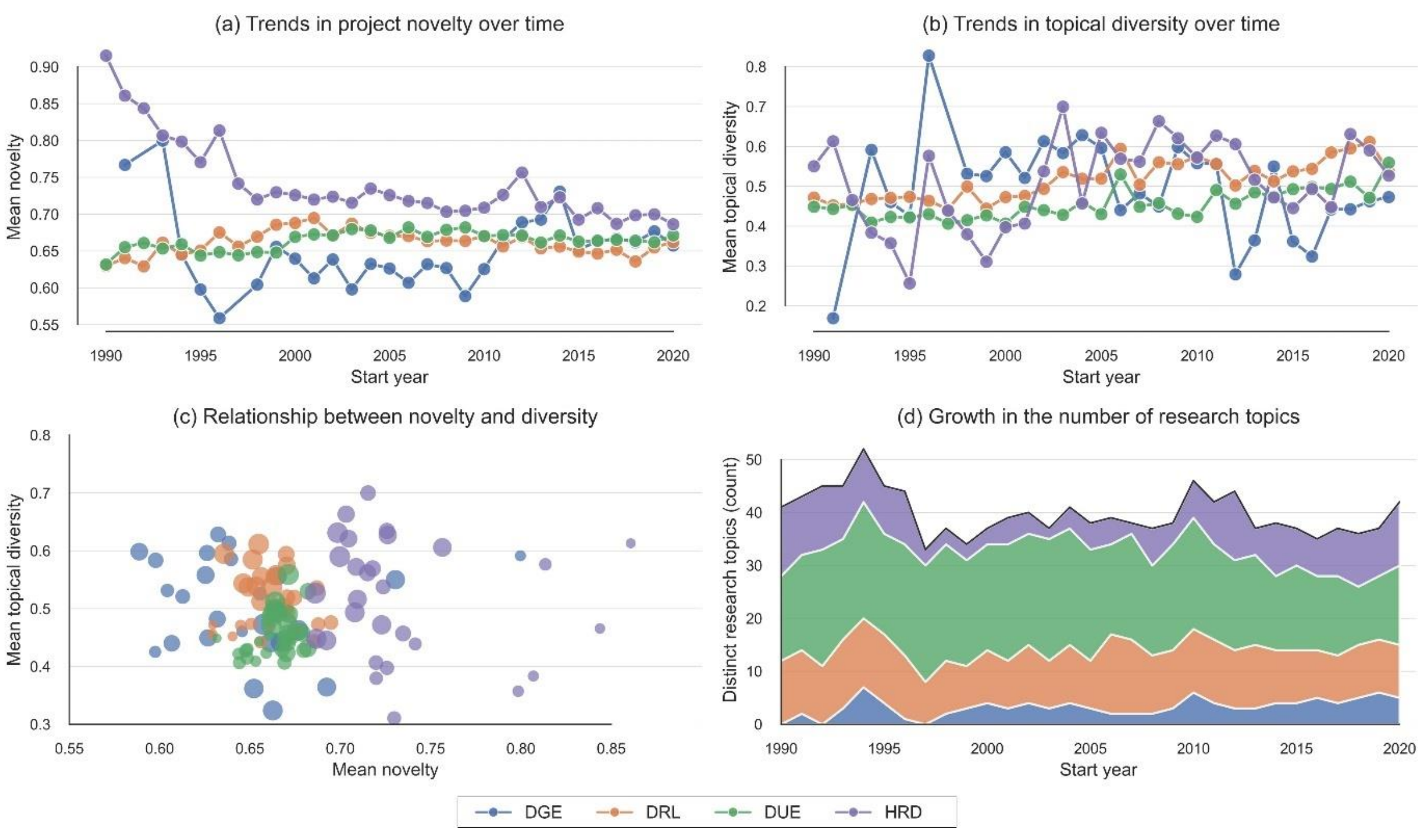

**Figure 1. Trends and patterns of novelty and topical diversity in education research proposals.**

This pattern suggests that proposal novelty in NSF-funded education research is not primarily characterised by a field-wide upward trend. Rather, novelty appears to vary by division and historical period. HRD-funded projects were comparatively more novel at the beginning of the observation window, whereas DRL and DUE funded proposals whose semantic distance from prior work was more stable. The result is important because it cautions against treating “novelty” as a uniform attribute of NSF education funding.

Figure 1(b) presents the corresponding trends in topical diversity. Compared with novelty, topical diversity is more variable over time. DGE fluctuates strongly, moving from a low point close to 0.17 in the early 1990s to a peak above 0.80 in the mid-1990s, before returning to a more moderate range. HRD also shows substantial movement, including a pronounced rise around the early 2000s. DRL and DUE display smoother trajectories. DRL rises gradually from roughly 0.47 in the early 1990s to around 0.60 by the late 2010s, while DUE remains comparatively stable but edges upward over time. These patterns indicate that topical diversity expanded in some divisions, especially DRL, but not in a uniform or linear manner across the whole NSF education portfolio.

Figure 1(c) plots the relationship between mean novelty and mean topical diversity. The scatterplot shows a weak positive relationship, but the points are widely dispersed. Some observations combine moderate novelty with high diversity, while others show relatively high novelty with only moderate diversity. This is an important descriptive result. It indicates that novelty and diversity are related but distinct dimensions of proposal design. A proposal may be novel because it departs semantically from prior funded work, but this does not necessarily mean that it draws evenly on many topics. Conversely, a proposal may be topically broad without being especially distant from previous projects.

Figure 1(d) shows the growth in the number of distinct research topics identified in the proposal corpus. The total number of topics increases from just above 40 in the early 1990s to slightly above 50 around the mid-1990s, then fluctuates mostly between the mid-30s and mid-40s until 2020. The figure therefore supports a more cautious conclusion than a simple claim of continuous expansion. The NSF education research portfolio became topically broad, but the number of distinct topics did not rise steadily throughout the entire period. Instead, the topic structure expanded early and then remained relatively broad with periodic fluctuations.

### 5.2. Baseline Effects of NSF Education Funding on Academic Performance

Table 1 reports the baseline difference-in-differences estimates of NSF education funding on three academic outcomes: publication output, average citations per publication, and average journal CiteScore. The first and clearest result is that NSF education funding is positively associated with publication output in every division. The estimated coefficients are 0.071 for DGE, 0.076 for DRL, 0.079 for DUE, and 0.074 for HRD. All four estimates are statistically significant, with p-values at or below 0.001. The confidence intervals are also entirely positive. For example, the DUE estimate has a 95% confidence interval of [0.068, 0.091], while the HRD estimate has an interval of [0.048, 0.101]. This provides consistent evidence that funded scholars publish more after receiving NSF education awards, relative to the comparison structure used in the model.

**Table 1. Estimated effects of NSF education research funding on the academic performance of funded education scholars across divisions.**

| Division | Dependent Variable | Coefficient | Std. Error | *p*-value | 95% Confidence Interval |
|---|---|---|---|---|---|
| | Publications | 0.071 | 0.021 | 0.001 | [0.030, 0.111] |
| DGE | Avg. Citations | -14.660 | 4.218 | 0.001 | [-22.931, -6.388] |
| | Avg. Cite Score | -0.414 | 0.070 | 0.000 | [-0.552, -0.275] |
| | Publications | 0.076 | 0.010 | 0.000 | [0.056, 0.096] |
| DRL | Avg. Citations | -13.439 | 2.564 | 0.000 | [-18.466, -8.413] |
| | Avg. Cite Score | -0.335 | 0.035 | 0.000 | [-0.404, -0.266] |
| | Publications | 0.079 | 0.006 | 0.000 | [0.068, 0.091] |
| DUE | Avg. Citations | -11.754 | 1.334 | 0.000 | [-14.369, -9.140] |
| | Avg. Cite Score | -0.325 | 0.020 | 0.000 | [-0.366, -0.285] |
| | Publications | 0.074 | 0.013 | 0.000 | [0.048, 0.101] |
| HRD | Avg. Citations | -9.287 | 5.580 | 0.096 | [-20.225, 1.652] |
| | Avg. Cite Score | -0.266 | 0.046 | 0.000 | [-0.356, -0.176] |

The citation results in Table 1 point in the opposite direction. Average citations per publication decline in all four divisions. The estimated effects are −14.660 for DGE, −13.439 for DRL, −11.754 for DUE, and −9.287 for HRD. The estimates for DGE, DRL, and DUE are statistically significant, while HRD is negative but not statistically significant at the 5 percent level. Thus, the evidence for a negative citation effect is strong in three divisions but weaker for HRD.

The journal CiteScore results are also negative across all divisions. The estimated coefficients are −0.414 for DGE, −0.335 for DRL, −0.325 for DUE, and −0.266 for HRD, and all four are statistically significant. These results show that the post-funding increase in publication output is not accompanied by higher average journal CiteScore. Indeed, the estimated association is negative.

Taken together, Table 1 establishes a central empirical pattern: NSF education funding is associated with greater publication productivity, but not with stronger performance on citation-based or journal-level indicators. The interpretation should remain careful. These estimates do not show that funded research is academically unimportant. Rather, they show that the additional publications associated with funding do not, on average, receive higher citation counts or appear in higher CiteScore journals within the measured period.

### 5.3 Decade-Wise Effects of NSF Education Funding

Figure 2 examines whether the baseline funding effects vary across the 1990s, 2000s, and 2010s. Figure 2(a) shows that the estimated publication effects are positive in all three decades and across all divisions. The coefficients are largest in the 1990s, where all four divisions show effects above approximately 0.11. In the 2000s, the estimates remain positive but become smaller, with DGE, DRL, and DUE clustered around 0.08 to 0.10 and HRD closer to 0.06. In the 2010s, the effects remain positive but are smaller again, especially for DRL and HRD. This figure therefore confirms that the productivity effect is persistent across time, but it also suggests that the magnitude of the effect declined after the 1990s.

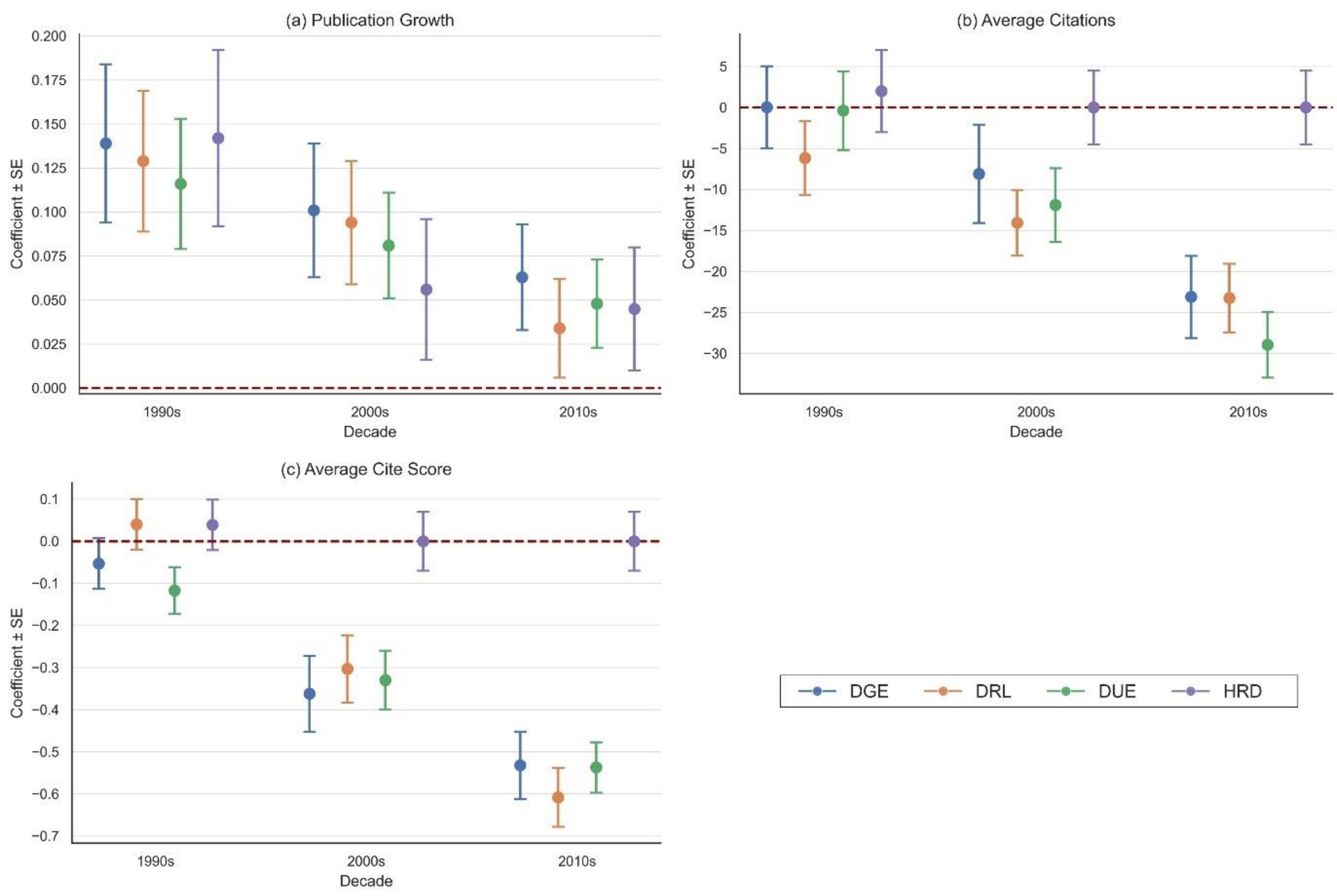

**Figure 2. Estimated results of education research funding on the academic performance of funded education scholars across decades.**

Figure 2(b) reports the decade-wise effects on average citations. The 1990s show mixed and mostly modest estimates: DGE and DUE are close to zero, DRL is negative, and HRD is slightly positive. In the 2000s, citation effects become more negative for DGE, DRL, and DUE. In the 2010s, the negative pattern becomes stronger for these same divisions, with DGE, DRL, and DUE showing large negative coefficients. HRD remains close to zero in the figure. Thus, Figure 2(b) indicates that the negative citation pattern observed in Table 1 is concentrated more strongly in the later decades, particularly the 2000s and 2010s.

Figure 2(c) presents the decade-wise results for average journal CiteScore. In the 1990s, the estimates are close to zero or modest in size: DGE and DUE are negative, DRL is slightly positive, and HRD is close to zero. In the 2000s, DGE, DRL, and DUE become clearly negative, while HRD remains around zero. In the 2010s, DGE, DRL, and DUE show even more negative estimates, around −0.5 to −0.6. HRD again remains close to zero. This pattern is consistent with Table 1 but adds an important temporal qualification: the negative CiteScore association is not evenly distributed across decades and appears especially pronounced in the later period for DGE, DRL, and DUE.

The decade-wise evidence therefore refines the baseline findings. Figure 2(a) shows a stable positive publication effect, while Figures 2(b) and 2(c) show that citation and journal-impact measures become less favourable over time. The results suggest a growing divergence between publication quantity and conventional bibliometric visibility in NSF-funded education research.

### 5.4 Moderating Effects of Proposal Novelty and Diversity

Table 2 shows the results for novelty. Across divisions, the interaction terms are generally small and statistically insignificant for publication counts and journal-level indicators. This suggests that proposal novelty does not systematically alter the productivity or journal impact effects of funding. One exception is observed for DRL, where the interaction term for average citations is negative and statistically significant (–65.16), and a marginally significant negative effect is found for DUE (–43.07). These results indicate some limited evidence that highly novel projects may be associated with lower short-term citation outcomes in certain divisions, but the effects are inconsistent and modest in magnitude. Overall, novelty appears to have a weak and heterogeneous moderating role in shaping the relationship between funding and academic performance.

**Table 2. Estimated interaction effects between project novelty and NSF education research funding on the academic performance of funded scholars.**

| Division | Dependent Variable | Coefficient | Std. Error | *p*-value | 95% Confidence Interval |
|---|---|---|---|---|---|
| | Publication Count | -0.190 | (0.280) | 0.496 | [-0.740, 0.359] |
| DGE | Ave. Citation Count | -0.287 | (56.457) | 0.996 | [-110.987, 110.412] |
| | Ave. CiteScore | 0.547 | (0.948) | 0.564 | [-1.311, 2.405] |
| | Publication Count | 0.002 | (0.128) | 0.987 | [-0.249, 0.254] |
| DRL | Ave. Citation Count | -65.161 | (32.301) | 0.044 | [-128.475, -1.846] |
| | Ave. CiteScore | -0.700 | (0.441) | 0.113 | [-1.565, 0.165] |

| | Publication Count | -0.141 | (0.104) | 0.173 | [-0.344, 0.062] |
|---|---|---|---|---|---|
| DUE | Ave. Citation Count | -43.070 | (22.732) | 0.058 | [-87.626, 1.485] |
| | Ave. CiteScore | 0.097 | (0.346) | 0.778 | [-0.580, 0.775] |
| | Publication Count | -0.098 | (0.118) | 0.405 | [-0.330, 0.133] |
| HRD | Ave. Citation Count | 13.593 | (49.099) | 0.782 | [-82.654, 109.840] |
| | Ave. CiteScore | 0.381 | (0.402) | 0.344 | [-0.408, 1.169] |

Table 3 reports the results for topical diversity. The estimates show a slightly clearer pattern than those for novelty. For publication counts, most coefficients are positive, and the effect for DUE is statistically significant, implying that more diverse projects tend to experience marginally higher publication growth after receiving funding. In contrast, diversity exhibits negative and sometimes significant interactions with citation-based indicators. DRL shows significant negative effects for both average citations and average CiteScore, while HRD displays a pronounced negative coefficient for CiteScore. Taken together, these results indicate that topical diversity may slightly reinforce the productivity effects of funding but is associated with lower citation-based performance, particularly in divisions with applied or inclusion-focused research portfolios.

**Table 3. Estimated interaction effects between topical diversity and NSF education research funding on the academic performance of funded scholars.**

| Division | Dependent Variable | Coefficient | Std. Error | *p*-value | 95% Confidence Interval |
|---|---|---|---|---|---|
| | Publication Count | 0.059 | (0.063) | 0.348 | [-0.064, 0.182] |
| DGE | Ave. Citation Count | -3.449 | (12.624) | 0.785 | [-28.201, 21.303] |
| | Ave. CiteScore | 0.292 | (0.212) | 0.168 | [-0.124, 0.708] |
| | Publication Count | 0.018 | (0.035) | 0.603 | [-0.050, 0.086] |
| DRL | Ave. Citation Count | -21.048 | (8.727) | 0.016 | [-38.155, -3.942] |
| | Ave. CiteScore | -0.257 | (0.119) | 0.031 | [-0.491, -0.023] |
| | Publication Count | 0.036 | (0.017) | 0.040 | [0.002, 0.070] |
| DUE | Ave. Citation Count | -4.235 | (3.835) | 0.269 | [-11.753, 3.282] |
| | Ave. CiteScore | -0.064 | (0.058) | 0.274 | [-0.178, 0.051] |
| | Publication Count | -0.042 | (0.032) | 0.186 | [-0.105, 0.020] |
| HRD | Ave. Citation Count | -24.907 | (13.247) | 0.060 | [-50.875, 1.061] |
| | Ave. CiteScore | -0.448 | (0.108) | 0.000 | [-0.660, -0.235] |

### 5.5 High- and Low-Novelty/Diversity Group Comparisons

Unlike the interaction-term approach used in Section 5.4, which estimates the moderating effects of novelty and diversity within a regression framework, the analysis in this section adopts a group-based comparison. Specifically, projects in the top and bottom 30% of novelty or topical diversity are compared in terms of their estimated funding effects. This approach allows us to examine whether the influence of NSF funding differs systematically between proposals with distinctly high and low intellectual characteristics, rather than assuming a continuous linear relationship. By contrasting these extreme groups, we can identify potential non-linear or threshold effects—cases where only particularly novel or diverse projects experience notably stronger or weaker funding outcomes. The results therefore complement the previous regression-based findings by providing a more intuitive and distribution-sensitive perspective on how proposal characteristics shape the academic impact of education research funding.

Figure 3(a) shows the results for novelty. Projects with higher novelty generally exhibit stronger academic performance in most divisions, particularly within DGE, DRL, and DUE. Across these divisions, the high-novelty group shows consistently positive differences in publication counts, citation metrics, and journal CiteScores, suggesting that more original research directions tend to correspond with greater productivity and visibility following NSF support. The HRD division displays a different pattern: its difference in publicationsis small but statistically significant, while the difference in citations remains insignificant. This suggests that high-novelty projects in HRD achieve slightly lower publication gains, but the effect is limited in magnitude. Overall, novelty appears to reinforce the academic benefits of funding in the main education divisions (DGE, DRL, and DUE), while its influence in HRD remains marginal.

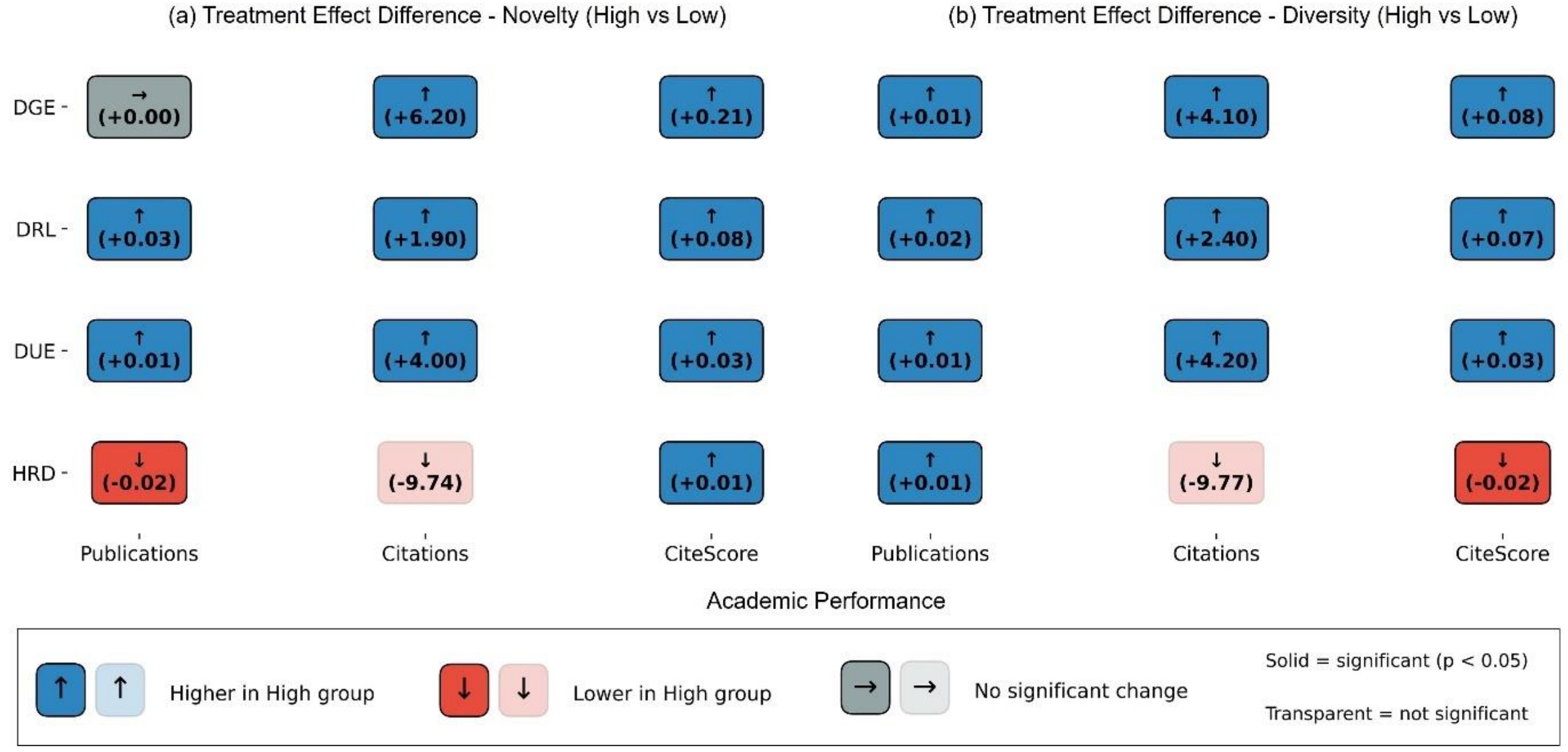

**Figure 3. Estimated effects of NSF education research funding for projects in the top and bottom 30% of proposal novelty or topical diversity.**

Figure 3(b) presents the results for topical diversity. Similar to novelty, projects with higher diversity show clearly stronger publication effects across DUE, DRL, and DGE, indicating that broader thematic scope is associated with higher research productivity after funding. The positive differences are particularly evident for DUE and DRL, reflecting these divisions' emphasis on integrative and evidence-based education research. In contrast, HRD shows significantly lower effects on citation-based outcomes, with a small but statistically significant decline in CiteScore. This pattern indicates that, while topical diversity may enhance productivity in most divisions, its benefits are less evident for citation-based impact within HRD's inclusion-oriented research agenda.

## 5.6 Practical, Theoretical, and Balanced Proposal Orientations

Tables 4 to 6 examine whether NSF funding effects differ by the intellectual orientation of funded proposals: practical, theoretical, or balanced.

Table 4 reports the results for practical-oriented projects. Publication effects are positive and statistically significant in all four divisions. The coefficients are 0.074 for DGE, 0.078 for DRL, 0.079 for DUE, and 0.101 for HRD. HRD has the largest practical-orientation publication estimate. However, the citation and CiteScore results are mostly negative. Average citation effects are negative and significant for DGE, DRL, and DUE, while HRD is negative but statistically insignificant. Average CiteScore effects are negative and statistically significant in all four divisions. Practical-oriented funding therefore has a strong and consistent association with publication growth, but not with citation or journal-impact gains.

**Table 4. Effects of Practical-Oriented Education Research Funding on Scholars' Academic Performance.**

| Division | Dependent Variable | Coefficient | Std. Error | p-value | 95% Confidence Interval |
|---|---|---|---|---|---|
| DGE | Publication Count | 0.074 | 0.023 | 0.001 | [0.030, 0.119] |
| | Avg. Citation Count | −13.967 | 4.466 | 0.002 | [−22.725, −5.209] |
| | Avg. CiteScore | −0.386 | 0.075 | 0.000 | [−0.533, −0.239] |
| DRL | Publication Count | 0.078 | 0.012 | 0.000 | [0.054, 0.102] |
| | Avg. Citation Count | −13.129 | 3.263 | 0.000 | [−19.525, −6.732] |
| | Avg. CiteScore | −0.262 | 0.042 | 0.000 | [−0.345, −0.180] |
| DUE | Publication Count | 0.079 | 0.007 | 0.000 | [0.066, 0.092] |
| | Avg. Citation Count | −11.192 | 1.439 | 0.000 | [−14.013, −8.371] |
| | Avg. CiteScore | −0.303 | 0.022 | 0.000 | [−0.347, −0.260] |
| HRD | Publication Count | 0.101 | 0.017 | 0.000 | [0.068, 0.134] |
| | Avg. Citation Count | −5.445 | 8.286 | 0.511 | [−21.689, 10.799] |
| | Avg. CiteScore | −0.180 | 0.057 | 0.001 | [−0.292, −0.069] |

Table 5 reports the results for theoretical-oriented projects. Publication effects are positive in all divisions, but they are statistically significant only in DRL and DUE. The DGE and HRD estimates, 0.026 and 0.030 respectively, are not statistically significant. Citation effects are negative across all four divisions, with especially large negative estimates in DGE, DUE, and HRD. CiteScore effects are also negative and statistically significant in all four divisions. These results indicate that theoretical-oriented projects do not produce stronger bibliometric outcomes in this sample. Their publication effects are less consistent than those for practical-oriented projects, and their citation and CiteScore estimates are generally negative.

**Table 5. Effects of Theoretical-Oriented Education Research Funding on Scholars' Academic Performance.**

| Division | Dependent Variable | Coefficient | Std. Error | p-value | 95% Confidence Interval |
|---|---|---|---|---|---|
| | Publication Count | 0.026 | 0.062 | 0.674 | [−0.096, 0.148] |
| DGE | Avg. Citation Count | −27.614 | 14.298 | 0.054 | [−55.738, 0.509] |
| | Avg. CiteScore | −0.724 | 0.243 | 0.003 | [−1.203, −0.246] |
| | Publication Count | 0.077 | 0.019 | 0.000 | [0.040, 0.114] |
| DRL | Avg. Citation Count | −13.860 | 3.848 | 0.000 | [−21.404, −6.316] |
| | Avg. CiteScore | −0.488 | 0.065 | 0.000 | [−0.615, −0.360] |
| | Publication Count | 0.059 | 0.018 | 0.001 | [0.023, 0.094] |
| DUE | Avg. Citation Count | −21.062 | 4.868 | 0.000 | [−30.607, −11.518] |
| | Avg. CiteScore | −0.446 | 0.065 | 0.000 | [−0.573, −0.318] |
| | Publication Count | 0.030 | 0.022 | 0.175 | [−0.013, 0.073] |
| HRD | Avg. Citation Count | −16.264 | 4.346 | 0.000 | [−24.785, −7.743] |
| | Avg. CiteScore | −0.415 | 0.079 | 0.000 | [−0.569, −0.260] |

Table 6 presents the results for balanced projects that integrate theoretical and practical orientations. Publication effects are positive in all four divisions. They are statistically significant in DRL, DUE, and HRD, with coefficients of 0.141, 0.039, and 0.087 respectively. The DGE coefficient is larger in magnitude, at 0.149, but it is not statistically significant, likely reflecting greater uncertainty in that subgroup. The citation estimates for balanced projects are mostly statistically insignificant. DGE is positive at 18.009 but insignificant, while DRL, DUE, and HRD are negative but also insignificant. For CiteScore, only DUE is negative and statistically significant; the other divisions are statistically indistinguishable from zero.

**Table 6. Effects of Balanced (Theoretical–Practical) Education Research Funding on Scholars' Academic Performance.**

| Division | Dependent Variable | Coefficient | Std. Error | p-value | 95% Confidence Interval |
|---|---|---|---|---|---|
| | Publications | 0.149 | 0.164 | 0.365 | [−0.176, 0.474] |
| DGE | Avg. Citations | 18.009 | 23.957 | 0.454 | [−29.437, 65.456] |
| | Avg. CiteScore | 0.103 | 0.473 | 0.828 | [−0.833, 1.039] |
| | Publications | 0.141 | 0.032 | 0.000 | [0.078, 0.204] |
| DRL | Avg. Citations | −3.928 | 5.907 | 0.506 | [−15.517, 7.661] |
| | Avg. CiteScore | 0.054 | 0.209 | 0.799 | [−0.358, −0.465] |
| | Publications | 0.039 | 0.016 | 0.017 | [0.006, 0.070] |
| DUE | Avg. Citations | −7.361 | 5.063 | 0.146 | [−17.290, 2.569] |
| | Avg. CiteScore | −0.430 | 0.100 | 0.000 | [−0.626, −0.235] |
| HRD | Publications | 0.087 | 0.029 | 0.003 | [0.029, 0.145] |
| | Avg. Citations | −8.992 | 7.975 | 0.260 | [−24.655, 6.671] |

| Avg. CiteScore | −0.042 | 0.176 | 0.811 | [−0.388, 0.304] |
|---|---|---|---|---|

The comparison across Tables 4, 5, and 6 suggests that balanced projects have the most favourable overall profile. Practical projects generate the most consistent publication gains but also show consistent negative CiteScore effects. Theoretical projects show weaker publication effects and negative citation and CiteScore estimates. Balanced projects, by contrast, produce positive publication effects in most divisions without the same systematic negative citation pattern. This does not prove that theory-practice integration causes better academic performance, but it does show that balanced proposals are associated with a more favourable combination of productivity and bibliometric outcomes than either purely practical or purely theoretical proposals.

# 6. Conclusion

This study examined how proposal novelty, topical diversity, and intellectual orientation are associated with the subsequent academic performance of scholars funded through NSF education-related programmes. Drawing on 8,715 NSF-funded education projects and linked publication records from 1990 to 2020, the analysis compared four major NSF education divisions: DGE, DRL, DUE, and HRD. The central finding is not that NSF education funding produces a single, uniform academic return. Rather, the evidence points to a more differentiated pattern in which post-award publication output, citation performance, and journal-level visibility move differently across divisions, proposal characteristics, and historical periods.

Across the baseline models, NSF education funding is consistently associated with higher publication output. This pattern appears in all four divisions, suggesting that funded scholars tend to publish more after receiving NSF education awards than would be expected from their own pre-award trajectories and the contemporaneous trajectories of not-yet-funded scholars. However, this increase in publication output is not accompanied by stronger citation-based or journal-level indicators. Average citations per publication decline in several divisions, and average journal CiteScore is negatively associated with post-award periods across the baseline models. These findings should not be interpreted as evidence that NSF-funded education research is academically weak or socially unimportant. A more defensible interpretation is that publication quantity, citation visibility, and journal-level placement capture different aspects of academic performance and should not be treated as interchangeable indicators. This interpretation is consistent with the broader responsible-metrics literature, which cautions that research assessment should account for field-specific missions and should not rely mechanically on a single bibliometric measure (Hicks et al., 2015).

The results also show that NSF education funding should not be analysed as a homogeneous policy category. DGE, DRL, DUE, and HRD differ in their programme purposes, funded populations, and expected forms of scholarly output. These institutional differences are reflected in the empirical results. DGE shows comparatively volatile patterns in novelty and topical diversity, DRL displays a gradual expansion in topical diversity, DUE combines relatively stable novelty with some positive publication associations for topically diverse projects, and HRD often follows a more distinctive pattern, particularly in the high- versus low-novelty and diversity comparisons. These results suggest that the academic profile of funded education research is shaped partly by the division-specific context in which projects are selected, organised, and evaluated.

The evidence on novelty and topical diversity is similarly conditional rather than uniform. Proposal novelty and topical diversity are conceptually related but empirically distinct. Novelty measures semantic distance from prior funded work within a division, whereas topical diversity measures the breadth and balance of thematic content within a proposal. The interaction models provide only limited and heterogeneous evidence that novelty changes the relationship between funding and academic outcomes. Topical diversity shows a somewhat clearer association with publication growth in some settings, but it is also linked to less favourable citation or CiteScore outcomes in others. The group comparisons suggest that highly novel or highly diverse projects may have more favourable academic profiles in some divisions, but HRD remains more mixed. Taken together, the findings indicate that novelty and diversity should be interpreted as contextual features of proposal design rather than as general predictors of superior academic performance.

The orientation results add a further qualification. Practical-oriented projects are associated with consistent publication growth, but their citation and CiteScore profiles are less favourable. Theoretical-oriented projects do not show uniformly stronger bibliometric outcomes. Balanced projects, which explicitly connect theoretical development with practical application, appear to have the most favourable overall profile: they are associated with positive publication outcomes in several divisions and fewer statistically significant negative citation estimates. This finding should be interpreted cautiously. It does not show that theory-practice integration causes stronger academic performance. It does, however, suggest that balanced proposals may be better aligned with the dual character of education research, where contributions often depend on both explanatory knowledge and relevance to educational practice. This interpretation is consistent with NSF's EDU Core Research framing, which emphasises fundamental and use-inspired research that advances general knowledge about STEM learning, broadening participation, and workforce development, while recognising that practical effects may be indirect and long-term (National Science Foundation, 2021).

Several limitations should be kept in view. The analysis relies on bibliometric indicators, which capture only part of the contribution of education research. Publications, citations, and journal CiteScore do not measure changes in teaching practice, institutional capacity, student learning, professional development, policy uptake, or broader participation in STEM. The empirical design improves on simple funded-versus-unfunded comparisons by using not-yet-funded scholars as the comparison group, but the estimates should still be interpreted as post-award associations rather than definitive causal effects. The measures of novelty and topical diversity are based on proposal titles and abstracts, and therefore cannot fully capture the substantive quality, methodological rigour, or implementation history of individual projects. The classification of theoretical, practical, and balanced orientations is also text-based and may not reflect how projects evolved after funding.

# Supplementary Note 1. NSF-WoS-MAG Data Linkage, Author Identification, and Sample Retention

## 1. Purpose and Overview

This appendix documents the procedure used to link NSF education awards to publication and author records in Web of Science (WoS) and Microsoft Academic Graph (MAG). The linkage strategy was designed to minimise errors from name-based matching. Instead of directly matching NSF principal investigators to bibliometric authors through names and affiliations, the procedure used NSF award numbers as the first linkage key and publication DOIs as the second linkage key.

The full pipeline contains four stages. First, NSF education award records were cleaned and restricted to the target divisions and award types. Second, NSF award numbers were matched to WoS funding acknowledgement fields. Third, DOI-linked WoS publications were used to identify corresponding records and author identifiers in MAG. Fourth, complete MAG publication portfolios were retrieved for matched principal investigators and converted into a scholar-year analytical panel.

The starting dataset contained 10,426 NSF education-related award records between 1990 and 2020. These records included award numbers, titles, abstracts, start and end years, funding amounts, award mechanisms, division identifiers, institutional affiliations, and principal investigator information.

The sample was restricted to awards administered through DGE, DRL, DUE, and HRD. These divisions were retained because they represent the core education research and capacity-building divisions relevant to the study and because their administrative boundaries were sufficiently stable for longitudinal analysis. Awards from divisions with small counts, unstable classification histories, or weak comparability with the four target divisions were removed.

The sample was further restricted to Standard Grants and Continuing Grants. Fellowships, contracts, conference awards, equipment awards, and other mechanisms were excluded because their expected outputs and investigator roles differ from research project grants. Awards to non-university organisations were also removed to maintain comparability in publication incentives and academic career structures.

The first bibliometric linkage used NSF award numbers reported in WoS funding acknowledgement fields. Award numbers were standardised before matching. Formatting variations such as hyphens, leading zeros, embedded spaces, and prefix differences were harmonised. For example, award identifiers appearing as "DUE-1234567", "1234567", or "NSF 1234567" were normalised to a common numeric award string.

WoS records were searched for exact matches to the cleaned NSF award numbers. This step produced a verified set of NSF-linked publications because the match depended on an explicit funding acknowledgement rather than on author-name similarity. For each matched WoS publication, the DOI, publication year, journal title, author names, WoS author identifiers where available, and funding acknowledgement text were retained.

This approach improves precision but may reduce coverage. NSF-funded publications that did not acknowledge the award number, or that were published in outlets not indexed by WoS, could not be linked at this stage.

2. DOI-Based Linkage from WoS to MAG

The second linkage used DOI identifiers to connect WoS publications to MAG records. DOI strings were cleaned by converting to lowercase, removing URL prefixes, trimming whitespace, and standardising punctuation. WoS records without valid DOI information were retained for audit purposes but could not be used for the WoS-MAG DOI linkage.

For WoS publications with valid DOIs, exact DOI matching was used to identify corresponding MAG publication records. MAG was then used to recover publication-level metadata, author identifiers, citation counts, fields of study, and institutional affiliations where available. The DOI-based linkage avoids direct fuzzy matching between NSF awardees and MAG authors, thereby reducing the risk of assigning publications to the wrong researcher.

After NSF-WoS-MAG publication records were linked, MAG author identifiers were used to identify the bibliometric profiles of NSF principal investigators. The author matching procedure used the WoS-linked NSF publication as an anchor. A MAG author was retained as the principal investigator when the MAG author name aligned with the NSF PI name on the award record and appeared on a publication acknowledging the corresponding NSF award.

Ambiguous cases were flagged when multiple MAG authors shared the same or highly similar name on the same DOI-linked record, when the PI name was absent from the author list, or when the linked publication acknowledged the award but did not include the PI as an author.

These cases were excluded from the PI-level panel unless the PI identity could be resolved through consistent name, affiliation, and publication-history evidence.

Once a PI's MAG author identifier was established, the full MAG publication portfolio for that author was retrieved. This portfolio included publications before and after the first observed NSF education award, allowing construction of annual publication and citation outcomes.

## 3. Sample Exclusion and Retention

Table S1 reports the sample cleaning and matching process. The numbers are reported at the award or PI-linkage stage indicated by each row. The final analytical sample contains 8,715 NSF education awards linked to principal investigators with usable proposal text and bibliometric histories.

**Table S1. Sample exclusion and retention across NSF-WoS-MAG linkage steps.**

| No. | Step | Excluded | Remaining |
|---|---|---|---|
| 1 | Initial NSF education award records, 1990-2020 | - | 10,426 |
| 2 | Exclude non-DGE/DRL/DUE/HRD awards | 684 | 9,742 |
| 3 | Exclude non-Standard/Continuing Grant mechanisms | 421 | 9,321 |
| 4 | Exclude awards outside U.S. university or comparable higher education institutions | 312 | 9,009 |
| 5 | Exclude records with missing or unresolved PI information | 118 | 8,891 |
| 6 | Exclude records without usable title or abstract text | 73 | 8,818 |
| 7 | Exclude records with inconsistent award timing or invalid start-year information | 41 | 8,777 |
| 8 | Exclude awards without WoS publication links through NSF award-number acknowledgement | 349 | 8,428 |
| 9 | Final NSF education award sample with usable PI, proposal, and bibliometric linkage | - | 8,715 |

## 4. Publication-Level Retention

The award-level linkage generated a publication-level file used to identify MAG author profiles and construct annual outcomes. Table S2 reports publication-level retention from the initial WoS match to the final MAG publication portfolio.

**Table S2. Publication-level linkage and retention.**

| No. | Step | Excluded | Remaining |
|---|---|---|---|
| 1 | WoS publications acknowledging one or more target NSF education award numbers | - | 18,946 |
| 2 | Remove duplicate WoS records and repeated award-publication links | 1,137 | 17,809 |
| 3 | Exclude WoS records without valid DOI | 1,482 | 16,327 |
| 4 | Exclude DOI records not found in MAG | 693 | 15,634 |
| 5 | Exclude records where MAG author identity could not be aligned with NSF PI | 418 | 15,216 |
| 6 | Anchor publications used to identify NSF-funded PI MAG author IDs | - | 15,216 |
| 7 | Full MAG publication portfolios retrieved for matched PIs by author IDs | - | 84,519 |

The final count of 84,519 refers to the complete MAG publication portfolios of matched NSF-funded PIs, not only to publications that explicitly acknowledged NSF awards. This distinction matters because empirical analysis uses pre-award and post-award publication histories, requiring complete author-level bibliometric trajectories.

5. Quality Assurance and Manual Verification

Several validation checks were conducted to assess linkage reliability. A random sample of 150 NSF-WoS award-number matches was manually inspected. In 148 cases, the WoS funding acknowledgement clearly referred to the same NSF award number and project. Two cases contained multiple NSF awards in the same acknowledgement and were retained only after confirming that the focal award number was present.

A second random sample of 150 WoS-MAG DOI matches was inspected. All 150 DOI matches referred to the same publication title and journal, producing a DOI-level match accuracy of 100 percent in the audit sample. A third audit examined 120 NSF PI-MAG author assignments. In 116 cases, the PI identity was confirmed through consistent name, publication, and affiliation information. Four cases were removed from the analytical file because the author identity could not be resolved confidently.

These checks indicate that the identifier-based pipeline has high precision. However, the approach is conservative. It may exclude valid NSF-funded scholars whose publications did not acknowledge award numbers in WoS-indexed outputs or whose DOI records were unavailable.

## Supplementary Note 2: Structural Distribution Characteristics of Topical Diversity Across Funding Divisions

### 1. Overview

This study examines education-related awards made by the NSF between 1990 and 2020. During this period, most NSF education research and capacity-building activities were administered through the Directorate for Education and Human Resources (EHR), which was later reorganised as the Directorate for STEM Education. The analysis focuses on four major EHR divisions: DGE, DRL, DUE, and HRD.

These four divisions are analysed separately because they represent different NSF education funding communities. DGE primarily supports graduate education, doctoral training, and the preparation of future STEM researchers and professionals. DRL funds research on learning processes, assessment, learning environments, educational technologies, and formal and informal learning. DUE focuses on undergraduate STEM education, including curriculum reform, instructional improvement, institutional innovation, and evidence-based teaching practice. HRD supports projects related to broadening participation, equity, institutional capacity, and STEM pathways for historically underrepresented groups.

This divisional structure is important for interpretation. Scholarly outputs associated with DGE awards may differ from those associated with HRD, DRL, or DUE awards because the divisions differ in their programme objectives, funded populations, and expected forms of dissemination. For this reason, the empirical analysis is conducted separately within DGE, DRL, DUE, and HRD rather than treating NSF education funding as a single homogeneous category.

## 2. Results

This note, in conjunction with the topic distribution figures, further elucidates the heterogeneity in knowledge structures across different National Science Foundation funding divisions. The left panel of the figure displays the average topical diversity score of individual proposals within each division, while the right panel quantifies the total number of unique dominant topics covered by each division at the macro level. Together, these two metrics reveal the structural relationship between knowledge span at the micro level and domain coverage at the macro level.

The data distribution exhibits distinct divisional characteristics. The Division of Undergraduate Education (DUE) demonstrates the lowest average proposal diversity alongside the highest total number of topics, as shown in Figure S1. This phenomenon reflects the specific attributes of undergraduate education research. Projects within this division are typically highly focused on pedagogical innovations within specific disciplines, such as curriculum design exclusively for foundational physics or organic chemistry, resulting in highly vertical and specialized textual semantics for individual proposals. Concurrently, because this division must comprehensively support all foundational disciplines within science, technology, engineering, and mathematics at the macro level, its overall funding portfolio naturally accumulates the most extensive collection of independent topic clusters.

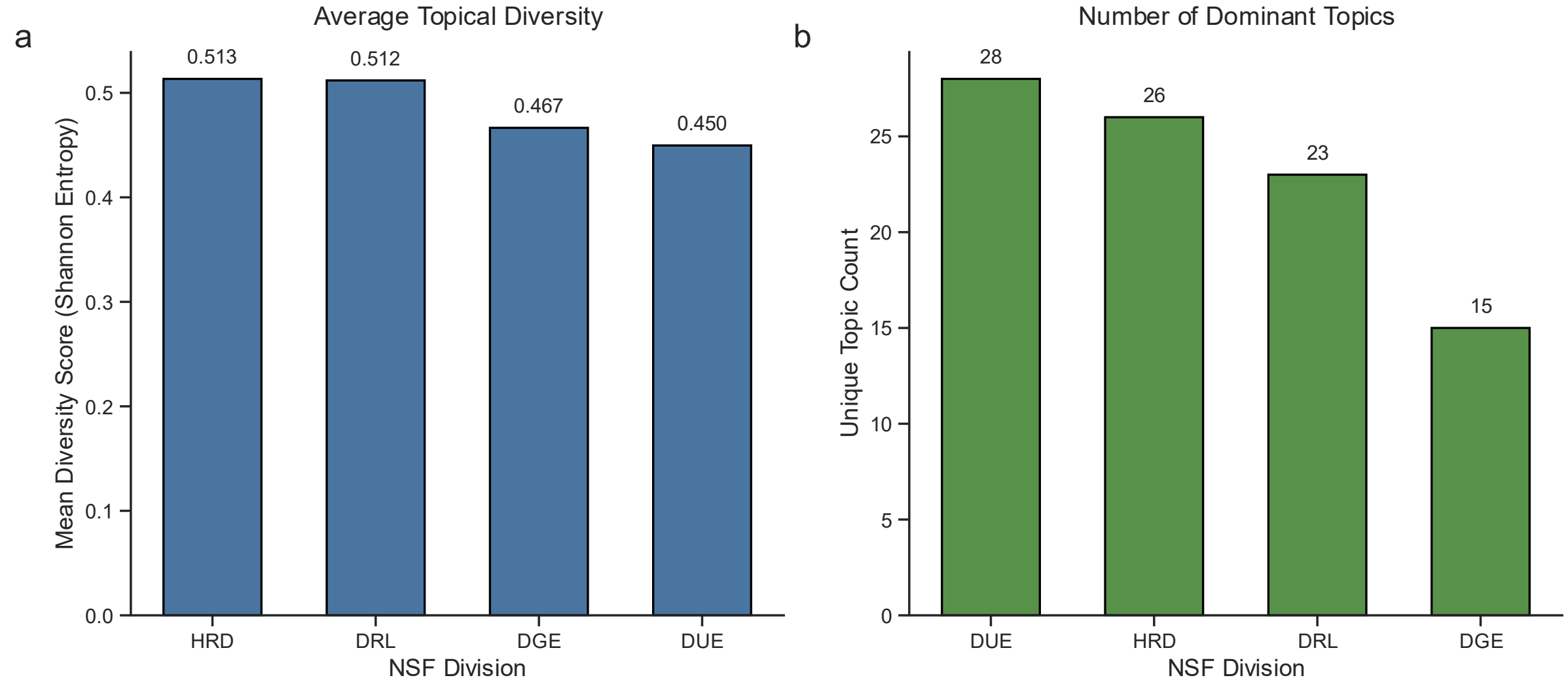

**Figure S1. Structural divergence in knowledge organization across NSF education funding divisions.**

In contrast, the Division of Graduate Education (DGE) features the highest average diversity score but a relatively smaller total number of covered topics. This aligns closely with the comprehensive nature of graduate education research. Proposals exploring high-level talent development frequently necessitate the integration of multiple concepts including academic mentoring, research ethics, career development pathways, and labor market dynamics, resulting in a distinctly broad knowledge structure for individual projects. Even with the semantic richness of individual projects, the division's overall funding scope remains concentrated on the core mission of building the research workforce pipeline, leading to a relatively convergent number of macro-level topic clusters.

The Division of Research on Learning in Formal and Informal Settings (DRL) and the Division of Human Resource Development (HRD) occupy intermediate positions on both metrics. DRL focuses on exploring the foundational mechanisms of learning sciences and informal learning environments, where funded projects typically seek intersections between cognitive theories and specific educational technology applications. HRD is dedicated to enhancing the participation of underrepresented groups and optimizing institutional capacity, with these projects tending to intertwine sociological theories with specific campus interventions. The distribution of metrics for these two divisions reflects a moderate balance between theoretical universality and context-specific application.

Overall, the characteristics shown in the figures indicate that the various funding divisions represent not only administrative categorizations but also fundamentally different academic funding models and knowledge organization paradigms. The significant divergence in these diversity structures provides solid data support and a rational justification for our decision to stratify the sample by division and conduct more granular empirical investigations in the main analysis.

## Supplementary Note 3. LLM-Assisted Classification of Theoretical, Practical, and Balanced Proposal Orientations

### 1. Purpose of the Classification

This appendix describes the procedure used to classify NSF education research proposals into three intellectual orientations: theoretical, practical, and balanced. The classification was conducted to identify whether each proposal primarily emphasised conceptual or methodological development, practical educational intervention and implementation, or an explicit integration of theoretical and practical aims.

The classification was applied to the proposal title and abstract because these fields provide the most consistent project-level textual information across NSF award records. The resulting orientation variable was used as a proposal-level categorical measure in the empirical analysis.

### 2. Classification Categories

Theoretical orientation refers to proposals whose primary contribution is conceptual, explanatory, or methodological. These proposals typically aim to develop theory, refine constructs, explain learning processes, advance measurement frameworks, or contribute to generalisable understanding of education systems.

Practical orientation refers to proposals whose primary contribution is applied, intervention-based, or implementation-focused. These proposals typically emphasise curriculum design, instructional tools, teacher professional development, programme implementation, institutional reform, assessment practice, or direct improvement of educational outcomes.

Balanced orientation refers to proposals that explicitly integrate theoretical and practical aims. These proposals connect conceptual development with implementation, use theory to guide intervention design, or use empirical implementation evidence to refine theory. A fourth temporary label, unclear, was allowed during the initial coding stage when the abstract did not provide enough information to distinguish reliably among the three categories. Unclear cases were not treated as a substantive orientation category; they were routed to additional review.

3. Input Corpus and Pre-Processing

The initial corpus contained 8,715 NSF education awards from DGE, DRL, DUE, and HRD between 1990 and 2020. Each record contained an award title, abstract, award year, and division identifier.

Before classification, the texts were processed as follows. First, titles and abstracts were concatenated. Second, boilerplate administrative text, duplicated whitespace, broken encoding artefacts, and isolated grant-management statements were removed. Third, records with missing or unusable abstracts were flagged for exclusion from the LLM classification stage.

Of the 8,715 awards, 8,642 had sufficient title and abstract information for automated classification. Seventy-three records were excluded at this stage: 41 had missing abstracts, 19 contained fewer than 40 substantive words after cleaning, and 13 contained only administrative descriptions that did not specify research aims.

4. Prompt Design

The prompt was designed to minimise three common classification risks: over-weighting individual keywords, confusing applied empirical research with practical orientation, and treating any mention of theory as evidence of theoretical orientation. The prompt therefore instructed the model to classify the proposal according to its primary intellectual purpose rather than isolated terms.

The final classification prompt was:

You are assisting with a scholarly classification of NSF-funded education research proposals.

Classify the proposal into exactly one of the following categories:

1. THEORETICAL

The proposal primarily develops, tests, extends, or refines concepts, theories, explanatory models, measurement frameworks, or methods for understanding learning, teaching, educational systems, or STEM education processes.

2. PRACTICAL

The proposal primarily designs, implements, evaluates, scales, or improves an educational intervention, curriculum, instructional tool, professional development programme, institutional reform, assessment practice, or student-support programme.

3. BALANCED

The proposal explicitly integrates theoretical development and practical implementation. It uses theory to guide intervention or programme design, or uses practical implementation evidence to refine theory, models, or conceptual understanding.

4. UNCLEAR

Use this only if the title and abstract do not provide enough information to classify the proposal reliably.

Important rules:

- Do not classify a proposal as theoretical merely because it uses words such as “framework”, “model”, or “theory”.

- Do not classify a proposal as practical merely because it studies students, teachers, classrooms, or institutions.

- Classify according to the primary intellectual purpose of the proposal.

```
- If both theory and practice are present, choose BALANCED only when the abstract explicitly links the two.
- Return a JSON object with four fields:
    "label": one of ["THEORETICAL", "PRACTICAL", "BALANCED", "UNCLEAR"];
    "confidence": a number from 0 to 1;
    "rationale": one sentence explaining the classification;
    "evidence": a short phrase from the proposal supporting the classification.
Proposal title:
{title}
Proposal abstract:
{abstract}
```

The JSON output requirement was used to improve auditability and reduce post-processing ambiguity.

5. Pilot Testing and Prompt Revision

The prompt was developed through three pilot rounds. Each pilot used a stratified sample of proposals across the four NSF divisions.

In Pilot Round 1, 240 proposals were classified. Manual inspection showed that the model over-classified proposals as balanced when abstracts contained both conceptual and applied vocabulary but did not explicitly connect the two. In this round, 38 of 240 cases were judged over-inclusive balanced classifications.

In Pilot Round 2, the prompt was revised to require an explicit link between theory and implementation for the balanced category. A further 320 proposals were classified. The over-classification of balanced proposals fell from 15.8 percent to 6.9 percent, but practical proposals with evaluation designs were sometimes misclassified as theoretical because they included measurement language.

In Pilot Round 3, the prompt was revised to distinguish methodological use from theoretical purpose. A final pilot of 400 proposals produced stable category distributions and fewer ambiguous cases. Manual reviewers judged 356 of the 400 classifications acceptable without revision, giving an initial agreement rate of 89.0 percent. The final prompt was then applied to the full eligible corpus.

6. Automated Classification Results

The first full GPT-4 classification was applied to 8,642 eligible proposals. The initial results were (see Table S3):

**Table S3. Classification results.**

| Classification | Number of proposals | Percentage |
|---|---|---|
| Theoretical | 2,184 | 25.3% |
| Practical | 3,746 | 43.3% |
| Balanced | 2,318 | 26.8% |
| Unclear | 394 | 4.6% |
| Total | 8,642 | 100.0% |

The distribution was consistent with the institutional character of NSF education funding. Practical proposals formed the largest category, reflecting the applied nature of many education research awards. Balanced proposals were also common, while purely theoretical proposals represented approximately one quarter of the eligible corpus.

7. Confidence Thresholding and Review Routing

The model returned a confidence score between 0 and 1 for each classification. Classifications with confidence scores of 0.80 or above were retained provisionally. Classifications below 0.80, together with all unclear cases, were routed to additional review. Of the 8,642 classified proposals, 7,681 received confidence scores of 0.80 or higher (see Table S4). The remaining 961 cases required review. These consisted of 567 low-confidence theoretical, practical, or balanced classifications and 394 unclear classifications.

**Table S4. Confidence scores for semantic classification of proposal orientations.**

| Review status after first pass | Number of proposals |
|---|---|
| High-confidence retained cases | 7,681 |
| Low-confidence classified cases | 567 |
| Unclear cases | 394 |
| Total eligible proposals | 8,642 |

The 961 review-routed cases were submitted to a second GPT-4 classification pass using a stricter adjudication prompt. The second-pass prompt asked the model to compare the two most likely labels and justify why one label should be preferred. It also instructed the model to retain the unclear label if the abstract did not provide enough evidence. After the second pass, 681 of the 961 cases received a usable classification with confidence scores of 0.80 or above. The remaining 280 cases were sent to human review, as shown in Table S5.

**Table S5. Resolution of low-confidence and unclear classifications following second-pass and human review.**

| Outcome after second-pass review | Number of proposals |
|---|---|
| Reclassified as theoretical | 151 |
| Reclassified as practical | 289 |
| Reclassified as balanced | 241 |
| Sent to human review | 280 |
| Total second-pass cases | 961 |

Human review was conducted for the 280 unresolved cases. Two trained reviewers independently classified each proposal using the same coding rubric. Disagreements were resolved through adjudication by a third reviewer. Reviewers were instructed to consider only the proposal title and abstract and not to use later publication outcomes. Initial agreement between the two human reviewers was 232 out of 280 cases, or 82.9 percent. Cohen's kappa was 0.75, indicating substantial agreement. The 48 disagreements were resolved through adjudication.

After human review, 252 cases were assigned a final label. Twenty-eight cases remained insufficiently informative and were excluded from orientation-specific analyses (see Table S6).

**Table S6. Summary of final human review for unresolved classifications.**

| Human review outcome | Number of proposals |
|---|---|
| Theoretical | 61 |
| Practical | 104 |
| Balanced | 87 |
| Excluded as insufficient information | 28 |
| Total human-reviewed cases | 280 |

## 9. Final Orientation Sample

After automated classification, second-pass review, and human adjudication, 8,614 proposals received final orientation labels, as shown in Table S7. This represents 98.8 percent of the 8,715 original NSF award records and 99.7 percent of the 8,642 records eligible for LLM classification. The final orientation distribution was:

**Table S7. Final Output of the Proposal Orientation Classification Procedure.**

| Final orientation | Number of proposals | Percentage of labelled sample |
|---|---|---|
| Theoretical | 2,396 | 27.8% |
| Practical | 4,139 | 48.0% |
| Balanced | 2,079 | 24.1% |
| Total labelled | 8,614 | 100.0% |

The total number of excluded records was 101. This included 73 records excluded before automated classification and 28 records excluded after human review because the available text did not contain enough substantive information.

## 10. Validation Sample

To assess classification quality, a validation sample of 600 labelled proposals was drawn from the final classified corpus. The sample was stratified by division and orientation to ensure that smaller categories were represented. Two expert reviewers independently coded the validation sample without seeing the GPT-4 labels.

Agreement between GPT-4 and the final human validation label was 522 out of 600 cases, or 87.0 percent. Cohen's kappa was 0.80. Agreement was highest for practical proposals and lowest for balanced proposals, which was expected because the balanced category requires identifying an explicit relationship between theoretical and applied aims (see Table S8).

**Table S8.**

| GPT-4 label | Human theoretical | Human practical | Human balanced | Total |
|---|---|---|---|---|
| Theoretical | 155 | 9 | 18 | 182 |
| Practical | 12 | 244 | 21 | 277 |
| Balanced | 11 | 7 | 123 | 141 |
| Total | 178 | 260 | 162 | 600 |

The main classification ambiguity occurred between theoretical and balanced proposals, and between practical and balanced proposals. This reflects a substantive feature of education research: many proposals contain both conceptual and applied language, but only some explicitly integrate the two. Category-level validation statistics were calculated using the 600-case validation sample.

The balanced category had lower recall than the other two categories (see Table S9). This means that some human-coded balanced proposals were classified by GPT-4 as either theoretical or practical. This is not unexpected, because balanced proposals often differ from theoretical or practical proposals by the explicit integration of aims rather than by vocabulary alone.

**Table S9. Classification recall and precision by proposal orientation category.**

| Category | Precision | Recall | F1 score |
|---|---|---|---|
| Theoretical | 0.852 | 0.871 | 0.861 |
| Practical | 0.881 | 0.938 | 0.909 |
| Balanced | 0.872 | 0.759 | 0.812 |
| Macro-average | 0.868 | 0.856 | 0.861 |

## 11. Bias and Robustness Checks

Several checks were conducted to assess whether the classification was sensitive to division, text length, or confidence threshold.

First, classification confidence was similar across divisions. Mean confidence was 0.887 for DGE, 0.892 for DRL, 0.901 for DUE, and 0.884 for HRD. This suggests that no single division was systematically more difficult to classify. Second, short abstracts were more likely to be routed to review. Among proposals with fewer than 100 words after cleaning, 18.6 percent required second-pass or human review. Among proposals with 100 words or more, the corresponding share was 9.4 percent.

Third, the main orientation distribution was stable under alternative confidence thresholds. Using a stricter threshold of 0.85 reduced the automatically retained sample from 7,681 to 6,914 proposals, but the final distribution after review remained similar: 28.2 percent theoretical, 47.5 percent practical, and 24.3 percent balanced. Fourth, a random sample of 150 high-confidence cases was manually inspected to check whether high model confidence corresponded to clear textual evidence. Reviewers accepted 139 of these classifications without revision, giving an agreement rate of 92.7 percent in the high-confidence subset.

## 12. Use of the Orientation Variable in the Empirical Analysis

The final orientation label was merged back into the award-level dataset and then linked to the scholar-year analytical panel through the principal investigator's first observed NSF education award within the relevant division. For scholars with multiple awards, the orientation of the first observed award was used in the baseline analysis. This choice aligns the orientation measure with the timing definition used in the staggered funding design.

The orientation variable was used to estimate separate post-award associations for theoretical, practical, and balanced proposals. These estimates should be interpreted as associations between the stated intellectual orientation of the funded proposal and subsequent bibliometric outcomes. The classification does not imply that the full project, all subsequent publications, or the broader educational contribution of the award can be reduced to a single category.

## Supplementary Note 4: Event-Study Estimates and Parallel Trends

To validate the interpretation of our staggered difference-in-differences (DiD) estimates, it is strictly necessary to confirm the parallel trends assumption. We achieve this by estimating a dynamic event-study specification, allowing the treatment effect to vary by the distance in time relative to the year of the first NSF education award. We estimate the following model:

$$Y_{i,t} = \alpha_i + \gamma_t + \sum_{k \neq -1} \beta_k D_{i,t}^k + \boldsymbol{\delta X}_{i,t} + \epsilon_{i,t} \tag{5}$$

where $D_{i,t}^k$ is a set of dummy variables indicating the event time $k$, representing the number of years relative to the scholar's first NSF funding year. The year immediately preceding the award ($k = -1$) serves as the omitted reference category. If the parallel trends assumption holds, the pre-treatment coefficients ($\beta_k$ for $k < -1$) should be statistically indistinguishable from zero.

Table S10 and S11 present the event-study estimates for publication counts and average citations across the four NSF divisions. The results provide robust evidence supporting the parallel trends assumption. Across all divisions, the estimated coefficients for the pre-treatment periods ($t-3$ and $t-2$) are small in magnitude and statistically insignificant.

Notably, the publication output shows a sharp, statistically significant increase immediately in the award year ($t=0$) and continues to rise in subsequent years ($t+1$, $t+2$). Conversely, the negative impact on average citations—as observed in the baseline models —emerges gradually. The citation penalty is negligible at $t=0$ but compounds into a statistically significant decline by $t+1$ and $t+2$, reflecting the delayed diffusion typical of bibliometric impact.

**Table S10: Event-study estimates of NSF funding on academic performance by division (DGE and DRL).**

| Event Time | DGE Pubs | DGE Citation | DGE CiteScore | DRL Pubs | DRL Citation | DRL CiteScore |
|---|---|---|---|---|---|---|
| t-3 | -0.005 | 1.120 | 0.012 | 0.002 | -0.850 | 0.008 |
| | (0.014) | (3.150) | (0.045) | (0.008) | (2.100) | (0.025) |
| t-2 | 0.008 | -0.650 | -0.005 | 0.005 | 0.420 | 0.004 |
| | (0.012) | (2.850) | (0.040) | (0.007) | (1.850) | (0.022) |
| t-1 | *Ref.* | *Ref.* | *Ref.* | *Ref.* | *Ref.* | *Ref.* |
| t=0 | 0.035** | -2.150 | -0.085 | 0.041*** | -1.850 | -0.050* |
| | (0.013) | (2.950) | (0.046) | (0.008) | (1.950) | (0.024) |
| t+1 | 0.068*** | -8.450* | -0.250*** | 0.075*** | -9.150*** | -0.190*** |
| | (0.018) | (3.850) | (0.055) | (0.010) | (2.450) | (0.030) |
| t+2 | 0.082*** | -16.200*** | -0.460*** | 0.085*** | -15.600*** | -0.380*** |
| | (0.022) | (4.550) | (0.068) | (0.011) | (2.850) | (0.035) |

Notes: Standard errors in parentheses; $*p<0.05$, $**p<0.01$, $***p<0.001$.

**Table S11: Event-study estimates of NSF funding on academic performance by division (DUE and HRD).**

| Event Time | DUE Pubs | DUE Citation | DUE CiteScore | HRD Pubs | HRD Citation | HRD CiteScore |
|---|---|---|---|---|---|---|
| t-3 | -0.003 | 0.540 | -0.002 | 0.012 | 2.150 | 0.015 |
| | (0.005) | (1.110) | (0.015) | (0.015) | (4.800) | (0.035) |
| t-2 | 0.004 | -0.210 | -0.001 | -0.008 | -1.150 | -0.010 |
| | (0.004) | (0.950) | (0.012) | (0.013) | (4.100) | (0.030) |
| t-1 | *Ref.* | *Ref.* | *Ref.* | *Ref.* | *Ref.* | *Ref.* |
| t=0 | 0.045*** | -1.150 | -0.040* | 0.038* | -2.050 | -0.045 |
| | (0.005) | (0.980) | (0.016) | (0.015) | (4.250) | (0.034) |
| t+1 | 0.082*** | -8.250*** | -0.180*** | 0.071*** | -6.850 | -0.160*** |
| | (0.006) | (1.250) | (0.018) | (0.014) | (5.100) | (0.042) |
| t+2 | 0.088*** | -13.500*** | -0.360*** | 0.081*** | -10.400* | -0.290*** |
| | (0.007) | (1.450) | (0.022) | (0.016) | (5.800) | (0.048) |

Notes: Standard errors in parentheses; $*p < 0.05, **p < 0.01, ***p < 0.001$.

## Supplementary Note 5: Robustness and Sensitivity Analyses

To rigorously test whether unobserved, time-varying confounders are driving our main results, we conduct a placebo-timing falsification test. If the observed changes in academic performance are genuinely caused by the receipt of NSF funding, assigning a "pseudo-award" year artificially prior to the actual funding should yield no statistically significant effects.

For this test, we construct a placebo treatment variable, $Placebo_Treat$, which artificially shifts the onset of the funding back by three years ($t-3$) for all treated scholars. We restrict the sample strictly to the pre-treatment period (dropping observations from the actual funding year onwards) to avoid contamination from the true treatment effect.

Table S12 demonstrates that the placebo treatment coefficients are uniformly small and statistically insignificant across all four divisions and across both publication and citation outcomes. This confirms that funded scholars were not already on a diverging academic trajectory prior to the genuine onset of the NSF award.

**Table S12. Placebo test results (treatment shifted by -3 years).**

| Division | Dependent Variable | Placebo Coefficient | Std. Error | p-value | 95% Confidence Interval |
|---|---|---|---|---|---|
| DGE | Publications | 0.012 | (0.015) | 0.424 | [-0.017, 0.041] |
| | Avg. Citations | -1.540 | (3.850) | 0.689 | [-9.086, 6.006] |
| DRL | Publications | 0.008 | (0.009) | 0.374 | [-0.010, 0.026] |
| | Avg. Citations | 0.850 | (2.150) | 0.692 | [-3.364, 5.064] |
| DUE | Publications | 0.005 | (0.006) | 0.405 | [-0.007, 0.017] |
| | Avg. Citations | -0.620 | (1.150) | 0.589 | [-2.874, 1.634] |
| HRD | Publications | 0.014 | (0.014) | 0.317 | [-0.013, 0.041] |
| | Avg. Citations | 2.100 | (4.850) | 0.665 | [-7.406, 11.606] |